\documentstyle[prb,aps,floats,epsf]{revtex}
\begin{document}
\twocolumn[
\hsize\textwidth\columnwidth\hsize\csname@twocolumnfalse\endcsname
\draft    
\title{Evidence for magnetic pseudoscaling in overdoped 
La$_{\bf {2-x}}$Sr$_{\bf {x}}$CuO$_{\bf {4}}$}
\author{J. G. Naeini, X. K. Chen, K. C. Hewitt, and J. C. Irwin}
\address{Department of Physics, Simon Fraser University, Burnaby, 
British Columbia, V5A 1S6, Canada}
\author{T. P. Devereaux}
\address{Department of Physics, George Washington University, 
Washington DC20052, USA} 
\author{M. Okuya$^{\dagger}$, T. Kimura$^{\ddagger}$, and K. Kishio}
\address{Department of Superconductivity, University of Tokyo, 
Bunkyo-ku, Tokyo 113, Japan}
\date{Phys. Rev. B {\bf 57}, R11077, 1998}
\maketitle
\begin{abstract} 
We report the results of electronic Raman scattering experiments on an 
overdoped La$_{1.78}$Sr$_{0.22}$CuO$_{4}$ 
single crystal as a function of 
temperature. The scattering rate $\Gamma(\Omega \rightarrow 0$,T) has been 
determined from the normal state B$_{1g}$ spectra in the range 
50 K$\leq\,$T$\,\leq\,$300 K. $\Gamma$(T) decreases linearly from 300 K 
to about 175 K and then undergoes a reduction with respect to the expected 
mean-field behavior. This trend suggests a crossover to a magnetic
pseudoscaling regime at T$_{cr}\simeq160\,$K. The results are  
in good agreement with the prediction of the nearly antiferromagnetic Fermi 
liquid model. There is no evidence of a pseudogap in the spectra 
obtained from this overdoped sample. 
\end{abstract} 

\pacs{PACS numbers: 74.25.Gz, 74.72.Dn, 78.30.Er \hspace{1.9in}
cond-mat/9711272}
]
It is now quite clear that the normal state properties of high temperature 
superconductors are very different from those of a Fermi 
liquid (FL). For example in optimally doped materials the unusual nature 
of these properties are manifested in transport measurements 
\cite{allen,kishio,batlogg} by a resistivity that varies linearly with 
temperature and in Raman experiments \cite{bozovic} by a featureless 
electronic continuum that extends to large energies.  Recent studies 
\cite{warren,puchkov,loram1,tallon,harris,ding,hwang,chen1} of underdoped 
compounds have revealed even more remarkable deviations from FL behavior. 
Many experiments have provided evidence for a strong quasiparticle 
renormalization, or depletion of spectral weight, that sets in at a 
temperature T$^*>\,$T$_c$.  The results of most of these investigations 
have been interpreted in terms of the opening of a normal state pseudogap 
(PG), a term that is generally used to mean a large suppression of 
low energy spectral weight.  
Although a great deal of effort has been expended the physical origin of the 
PG remains unknown.  Recent photoemission experiments \cite{harris,ding} have 
found that the PG has the same symmetry as the superconducting gap. 
These results imply that the depression of spectral weight above T$_c$ is 
associated with precursor pairing.  For example it has been proposed 
\cite{emery,fisher}  that pair formation, without phase coherence, 
could occur at T$\,>\,$T$_c$ with phase coherence, and hence the transition 
to the superconducting state, being established at the lower temperature 
T$_c$. However, other experiments such as the specific heat measurements of 
Loram {\em et al}.\cite{loram2}, are consistent with a mechanism that 
competes with superconductivity for the available quasiparticles.  

Another issue concerns the nature of the normal state excitations in the 
overdoped regime. 
Although there is experimental evidence to suggest that the 
overdoped compounds behave more like normal metals there is considerable 
evidence to the contrary.  
Some workers \cite{batlogg,hwang,tatiana} have found that the PG is still 
present well into the overdoped state.  Transport measurements 
by Hwang {\em et al.} \cite{hwang} on overdoped  
La$_{\rm {2-x}}$Sr$_{\rm x}$CuO$_{\rm 4}$ [La214(x)] suggest that T$^*$ 
is much greater than T$_c$ for x$\,\leq\,$0.22.  More recently infrared 
reflectivity measurements \cite{tatiana} on La214(0.22) have been 
interpreted in terms of the onset of a PG at T$^*\approx\,$300 K.  
These measurements suggest that  the PG persists into the overdoped state 
and are consistent with a modified phase diagram proposed by Batlogg 
{\em et al.} \cite{batlogg} in which, for La214, T$^*$ is significantly 
greater than T$_c$ at optimum doping and becomes equal to T$_c$ well 
into the overdoped region.  

Recently Pines and coworkers \cite{pines1,pines2}, after analyzing 
NMR and neutron scattering experiments, have proposed a new phase 
diagram for the hole-doped cuprates that could provide an explanation for 
the apparently conflicting results described above. In their nearly 
antiferromagnetic Fermi liquid (NAFL) model a tendency to order 
antiferromagnetically competes for quasiparticles with a spin fluctuation 
mediated pairing mechanism.  They have discussed their results in terms of 
the variation of the anti-ferromagnetic (AF) correlation length $\xi$ with 
temperature and doping.  This leads to the definition of a temperature 
T$_{cr}$, at which $\xi($T$_{cr})\cong$ 2a, where a crossover occurs between 
what they define as a mean-field (MF) region (T$\,>\,$T$_{cr}$) to a region 
(T$^*<$\,T\,$\leq$T$_{cr}$) in which magnetic pseudo-scaling (PS) prevails 
and $\xi$ increases rapidly with decreasing temperature.  
At a lower temperature T$=$T$^*$ there is a crossover to the PG regime in 
which $\xi$ is approximately constant.  Both T$_{cr}$ and  T$^*$, and their 
difference, decrease with increasing doping and for
La214 approach T$_c$ well into the overdoped regime \cite{pines2}.  

To gain additional insight into the above issues concerning the normal state 
of overdoped cuprates, we have carried out electronic Raman scattering 
experiments on an overdoped La214(0.22) single crystal.  
In many ways La214 is a prototype material for these studies in that it 
has a single CuO$_2$ layer in the unit cell and this enables one to avoid 
any effects that might be associated with interlayer coupling.  
In addition, the overdoped state of La214 is well characterized and can be 
reproducibly attained \cite{kimura,keimer,radaelli}. Finally, La214 has been 
the subject of many of the previous studies cited above and thus our results 
can be compared directly to these studies and to the predictions of the 
model proposed by Pines and coworkers \cite{pines2}.  We have measured the 
low energy B$_{1g}$ and B$_{2g}$ Raman continua as a function of temperature. 
From each B$_{1g}$ spectrum we have obtained an estimate for the scattering 
rate $\Gamma(\Omega \rightarrow 0$,T). 
The dependence of $\Gamma$(T) on temperature is consistent with the 
existence of a crossover from MF to a PS behavior at T$_{cr}\simeq160\,$K.  
We have not found any evidence for a crossover to a PG regime , however, 
which implies that T$^*\leq\,$T$_c$ for this overdoped sample. 
The results are in general agreement with the predictions of the NAFL model.

The high-quality La214(0.22) single crystal (T$_c\,$=$\,$30K) 
used in this study was grown by a traveling solvent floating-zone method, 
as previously described \cite{kimura}. The specimen, with dimensions of 
$4 \times 2 \times 0.5$ mm$^3$, were oriented using Laue x-ray diffraction 
patterns. The surfaces of the sample were polished with diamond paste and
etched with a bromine-ethanol solution \cite{werder}.
Raman spectra were obtained in a quasibackscattering geometry using the 
514.5 nm line of Ar$^+$ laser, which was focused onto the sample with
a cylindrical lens to provide an excitation level of about 10 W/cm$^2$.
The temperature of the excited region of the sample was found to be about
$11\,$K above ambient. 
This has been estimated from the intensity ratio of the Stokes and 
anti-Stokes spectra. The temperatures reported in this paper are the 
actual temperatures of the excited region of the sample.

\begin{figure}[htb]
\centerline{\epsfxsize=2.8 in \epsffile{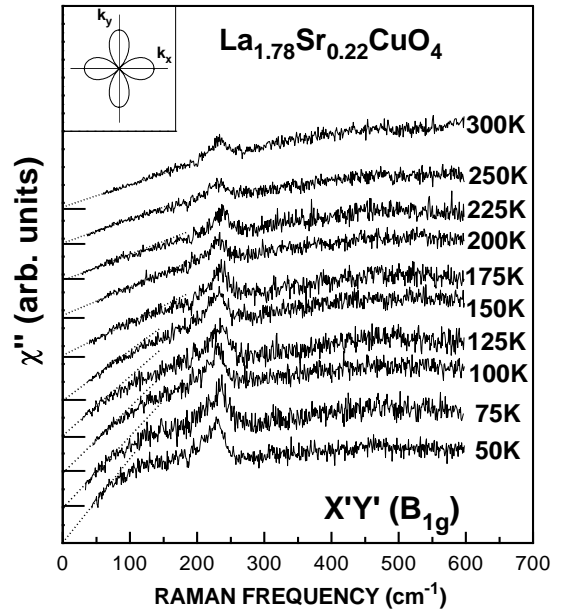}}
\vspace{0.1in}
\caption{The B$_{1g}$ Raman response ($\chi^{\prime\prime}_{B_{1g}}$) of La214(0.22) 
obtained at temperatures between $50\,$K and $300\,$K. The dashed lines are 
the slopes of $\chi^{\prime\prime}_{B_{1g}}$ as $\Omega \rightarrow 0$. 
The spectra have been displaced for clarity and their zeros are indicated 
by ticks on the vertical axis. The insert shows a polar plot of the 
angular dependence of the B$_{1g}$ component of the Raman tensor 
$\gamma_{x^{\prime}y^{\prime}}$.} 
\label{1}
\end{figure}
The scattering geometries used in this paper are defined by specifying the 
polarizations of the incident and scattered light with respect to a set of 
axes, $x(1,0,0)$ and $y(0,1,0)$, which are chosen to lie along the Cu-O bonds 
in the CuO$_2$ planes, or with respect to $x^{\prime}(1,1,0)$ and 
$y^{\prime}$(\={1},1,0) that are rotated 45 degrees with respect to $x$ 
and $y$. In all cases the incident and scattered light travels parallel to 
the $z(0,0,1)$ axis.  For non-resonant excitation the components of the Raman 
tensor are given approximately \cite{chen2,tom1} by $\gamma_{ij} 
\propto \partial^2\epsilon({\bf k})/\partial$k$_i \partial$k$_j$. 
The orthorhombic distortion is very small \cite{kimura,keimer,radaelli} 
for this overdoped crystal and thus tetragonal symmetry (D$_{4h}$ point group) 
can be used for a discussion of the polarization dependence of the Raman 
experiments. In this case $\gamma_{xy}$ must transform \cite{chen2} as 
k$_x$k$_y$ or the B$_{2g}$ irreducible representation of D$_{4h}$ and 
$\gamma_{x^{\prime}y^{\prime}}$ must transform as (k$^2_x-\,$k$^2_y$) 
or B$_{1g}$. Thus   $\gamma_{x^{\prime}y^{\prime}}$ has a maximum 
near the k$_x$ or k$_y$ axis and is zero along the diagonal directions, while 
$\gamma_{xy}$ will have the complementary symmetry dependence in k-space. 
In other words the $xy$ scattering geometry probes 
regions of the FS located near the diagonal directions  and the 
$x^{\prime}y^{\prime}$ spectra arise from excitations located near the 
$(1,0)$ axes.  These symmetry considerations can be illustrated by using a 
tight binding model to represent the band structure and then calculating 
\cite{chen1} the angular dependence of the components of the Raman tensor.
The result of such calculation carried out for 
$\gamma_{x^{\prime}y^{\prime}}$ is shown in the insert of Fig.~\ref{1}.

In an attempt to gain information on the scattering processes that influence
the B$_{1g}$ channel, we have studied the temperature-dependence of 
the B$_{1g}$ spectra. Fig.~\ref{1} shows the $x^{\prime}y^{\prime}$ 
Raman response of La214(0.22) obtained at temperatures between $50\,$K and 
$300\,$K. The Raman response functions for symmetry $\gamma$
($\chi^{\prime\prime}_{\gamma}$) are obtained 
by dividing the measured intensity by the thermal factor [1--exp(-$\hbar 
\Omega$/$k_B$T)]$^{-1}$. The dashed lines in the 
figure represent the slope of $\chi^{\prime\prime}_{\gamma}$ as 
$\Omega \rightarrow 0$. As shown in Fig.~\ref{1}, the low energy continua 
indicate a redistribution which gradually decreases with increasing  
temperature and changes trend for T$\,>150$ K. As will be described below, 
this behavior might be associated with the variation of 
the AF correlation length $\xi$ with temperature which suggests a 
crossover from MF to PS regime at T$_{cr}\simeq160$ K. Furthermore,
Fig.~\ref{1} indicates an increase in the intensity of the Raman
continuum near 500 cm$^{-1}$, once the temperature is decreased below
160 K. This behavior, as well as the higher frequency Raman response,
will be discussed in detail in a forthcoming publication.

\begin{figure}[htb]
\centerline{\epsfxsize=2.7 in \epsffile{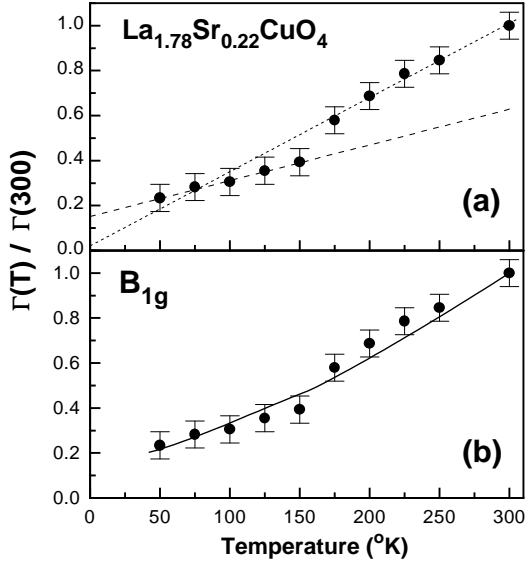}}
\vspace{0.1in}
\caption{(a) The normalized DC 
($\Omega \rightarrow 0$) scattering rate $\Gamma$(T)/$\Gamma$(300) 
in La214(0.22) obtained from the B$_{1g}$ spectra. The dashed lines are 
least squares linear fits to  MF and PS  regimes.
(b) Comparison of the normalized DC scattering rate with values (solid lines) 
calculated from the NAFL model.} 
\label{2}
\end{figure}

If both vertex corrections and the real part of the self energy are neglected,
an expression (Eq. 3 of Ref. 25) for the Raman response 
in the limit of $\Omega \rightarrow 0$ and T$\ll \mu$  can be written as 
\begin{equation}
\chi^{\prime\prime}_{\gamma}(\Omega,T) =\frac{2\Omega}{N} 
\sum_{\bf k}\gamma^{2}({\bf k})
{\Gamma^{2}({\bf k},T)\over{[(\epsilon({\bf k})-\mu)^{2}+
\Gamma^{2}({\bf k},T)]^{2}}}.
\label{chik}
\end{equation}
Here $\Gamma({\bf k},$T) is the momentum and temperature dependent scattering
rate which is equal to 
the imaginary part of the self energy evaluated on the FS,
$\epsilon({\bf k})$ is the band structure, $\mu$ is the chemical potential, 
and $N$ is the number of sites.
From Eq. (\ref{chik}) the slope of the low energy Raman response 
$\chi^{\prime\prime}_{\gamma}(\Omega \rightarrow 0$,T) is inversely 
proportional 
to $\Gamma({\bf k},$T) weighted by the Raman tensor $\gamma({\bf k})$
\begin{equation}
\left [\frac{\partial\chi^{\prime\prime}_{\gamma}(\Omega,T)}{\partial\Omega} 
\right ]_{\Omega \rightarrow 0} \propto \Gamma^{-1}_{\gamma}(T)=
\langle \gamma^{2}({\bf k})/\Gamma({\bf k},T)\rangle,
\label{dclim}
\end{equation}
where we have replaced the ${\bf k}$-sum with an integral over an infinite
band and an angular integral $\langle \cdots\rangle$ over the FS.
Using (\ref{dclim}) we can thus determine the B$_{1g}$ scattering rate from 
the inverse slope of  $\chi^{\prime\prime}_{B_{1g}}(\Omega \rightarrow 0$,T) 
shown in Fig.~\ref{1}.  
The results of these determinations could be seen in Fig.~\ref{2} where they 
are plotted in normalized form [$\Gamma$(T)/$\Gamma$(300)] as a function of 
temperature.  The variation of the scattering rate with temperature, as shown 
in Fig~\ref{2}(a), clearly suggests the identification of two distinct 
regimes with a crossover at approximately 160 K. These results appear to be 
in qualitative agreement with the predictions of NAFL model if we make 
the identification T$_{cr}\simeq160$ K.  This assignment is also in 
accord with the results of determination of T$_{cr}$ shown in Fig. 4 of 
Ref. \cite{pines1} obtained from transport and susceptibility measurements of 
La214.  
\begin{figure}[htb]
\centerline{\epsfxsize=2.9 in \epsffile{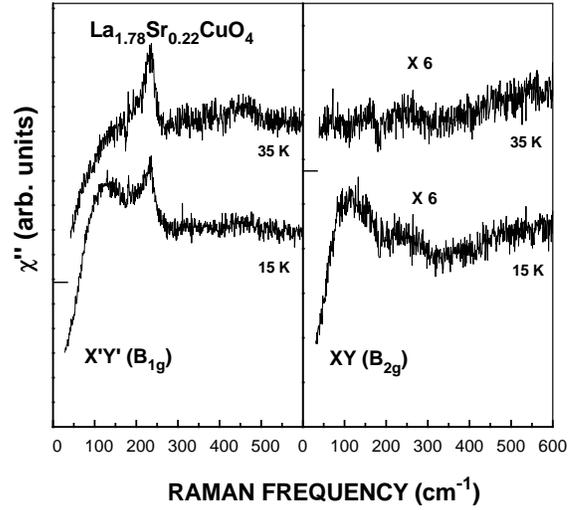}}
\vspace{0.1in}
\caption{The B$_{1g}$ and B$_{2g}$ Raman response functions 
($\chi^{\prime\prime}$) of La214(0.22) measured at 15 K and 35 K.
The B$_{1g}$ spectra are about 6 times stronger than the B$_{2g}$ spectra.} 
\label{3}
\end{figure}

These considerations 
are quite persuasive but to make the comparison more quantitative the 
Raman spectra for NAFL can be calculated \cite{tom2}. The results of the 
calculated scattering rates are compared to the measured values in 
Fig.~\ref{2}(b) where the solid lines indicate the results of evaluating 
Eq. (\ref{dclim}). Here we have used the electron-electron interaction 
\cite{pines1}
\begin{equation}
V({\bf q},\omega)=g^{2}{\alpha\xi^{2}\over{1+({\bf q-Q})^{2}\xi^{2}-i\omega/
\omega_{sf}}},
\label{nafl} 
\end{equation}
to calculate the self energy \cite{tom2}. In Eq. (\ref{nafl}) $g$ is the 
coupling constant,  $\omega_{sf}$ and $\xi$ are the phenomenological 
temperature dependent spin fluctuation energy scale and the correlation 
length, respectively, which can be determined via fits to magnetic 
response data\cite{pines1}. These functional parameters obey certain 
relations depending on different temperature and doping regimes. 
In the $z=1$ or PS regime, the spin correlations are strong enough 
to lead to changes from the classical MF theory $z=2$ regime. 
For each scaling regime, $\omega_{sf}\xi^{z}=$ constant (i.e. 
temperature independent). As in Ref. \cite{pines1} for
the $z=1$ scaling regime (T$^*<\,$T$\,\leq\,$T$_{cr}$)
we use $1/\xi=0.23+2.25\times10^{-3}T$[K] and further assume
that $\omega_{sf}\xi=72$meV for La214(0.22) with T$_{cr}=160\,$K. 
In the absence of any prediction for how these parameters crossover 
from one scaling regime to another, we assume a direct crossover to $z=2$ 
behavior given by $\omega_{sf}^{z=2}(T)/t=\omega_{sf}^{z=1}(T_{cr})/t+
0.4k_{B}(T- T_{cr})/t$, with $t=200$meV the near neighbor hopping,
such that for $z=2$ $\omega_{sf}\xi^{2}=122$meV and the correlation length 
and spin fluctuation energy smoothly cross over from $z=1$ to $z=2$ behavior 
at $T=T_{cr}$. In addition, we have used a tight binding energy band with the 
$t^{\prime}/t=0.16$ and filling $\langle n\rangle=0.8$ for both spins.
Lastly we add to Eq. (\ref{nafl}) an isotropic impurity interaction 
$H_{imp}=\sum_{\bf k,k^{\prime}}\sum_{i,\sigma}
U e^{i({\bf k-k^{\prime}})\cdot {\bf R}_{i}}
c^{\dagger}_{{\bf k},\sigma}c_{{\bf k^{\prime}},\sigma},$
where ${\bf R}_{i}$ denotes the position of the impurity labeled by
$i$ and $U$ is the impurity potential. After averaging over the position of 
the impurities, this adds a momentum independent term to the imaginary part 
of the self energy $\Gamma_{imp}=\pi n_{i}N_{F}\mid U\mid^{2}$, where 
$n_{i}$ is the impurity concentration, and $N_{F}$ is the density of states 
per spin at the Fermi level. To obtain the best fit to the data we have used 
$\Gamma_{imp}=9.7$cm$^{-1}$. The resultant agreement between theory and 
experiment is obviously quite satisfactory and makes a plausible case for the 
applicability of the NAFL. Further details are reported in Ref. \cite{tom2}.

According to Fig.~\ref{2}(a) in both scattering regimes $\Gamma$(T) varies 
linearly with temperature but with different slopes.
In the mean field region (T$\, >\,160\,$K) the electrons should 
be uncorrelated, and a linear  extrapolation to T = 0 should provide an 
estimate \cite{hackl} for the DC impurity scattering rate $\Gamma_{MF}(0)$.  
A high temperature extrapolation is shown in Fig.~\ref{2}(a) along with a 
linear extrapolation of the data for T$\, <\,160\,$K.  As is evident 
the extrapolation from the pseudoscaling regime yields 
$\Gamma_{PS}(0)\,>\,\Gamma_{MF}(0)$.  This increase in the estimate for 
residual scattering rate might be associated with scattering by 
collective spin fluctuations which takes place in the PS regime.  This 
suggestion is also consistent with the depletion of the low energy B$_{1g}$ 
spectrum  at temperatures just above T$_c$ (see Fig.~\ref{3}).  

To complete this discussion we note that the temperature dependence of the 
scattering rate does not provide any evidence for a crossover to a pseudogap
regime. Furthermore, in underdoped crystals the presence of the PG was 
associated \cite{chen1} with 
a significant depletion of spectral weight in the B$_{1g}$ channel as
is also found in photoemission experiments \cite{harris,ding}.  
This depletion resulted in the B$_{1g}$ spectrum being much weaker 
than the B$_{2g}$ spectrum.  However, the B$_{1g}$ spectrum of the 
overdoped crystal is very strong and in fact significantly more intense 
than the B$_{2g}$ spectrum as is shown in Fig.~\ref{3}.  
Finally, the presence of a PG tends to mask the transition to the 
superconducting state and hence there is no observable \cite{chen1} 
superconductivity induced renormalization in the B$_{1g}$ channel.  
However, as is clear from Fig.~\ref{3}, the B$_{1g}$ spectrum undergoes 
a very strong renormalization when the sample is cooled below T$_c$.  
These considerations strongly suggest that the PG is absent 
in La214(0.22).

In summary, we have carried out polarized Raman measurements of the low 
energy electronic continua of La$_{1.78}$Sr$_{0.22}$CuO$_{4}$ over a wide 
range of temperatures. From the low frequency B$_{1g}$ spectra we have 
estimated the temperature dependent scattering rate $\Gamma$(T) for 
quasiparticles on portions of the Fermi surface near ($\pm1,0$) and 
($0,\pm1$) directions. The results appear to be in good agreement with the 
predictions of the nearly antiferromagnetic Fermi liquid model and in 
particular a crossover from a mean-field to a pseudoscaling behavior 
at T$_{cr}\simeq160\,$K is suggested.  However, we have not found any 
evidence for a crossover to a pseudogap regime. 
This would imply that T$^*\leq\,$T$_c$ for this overdoped sample.

J.G.N. and J.C.I. gratefully acknowledge the financial support of 
the Natural Science and Engineering Council of Canada and  helpful discussions
with T. Startseva and T. Timusk. T.P.D. has benefited from helpful 
conversations with D. Pines and B. Stojkovic and thanks the Donors of the 
Petroleum Research Fund, administered by the American Chemical Society, for
partial support of this research.

{\small {$\dagger$Present address: Department of Material 
Science,Shizuoka University, Johoku, Hamamatsu 432, Japan.}}\\
{\small {$\ddagger$Present address: Joint Research Center for 
Atom Technology, Higashi, Tsukuba 305, Japan.}}


\begin{references}
\bibitem{allen} P. B. Allen, Z. Fisk, and A. Miglieri,  Physical
Properties of High Temperature Superconductors I, edited by D. M. Ginsberg
(World Scientific, 1989).
\bibitem{kishio} 
G. S. Boebinger {\it et al.},
Phys. Rev. Lett. {\bf 77}, 5417 (1996).
\bibitem{batlogg} B. Batlogg {\it et al.},
Physica C {\bf 235-240}, 130 (1994).
\bibitem{bozovic} I. Bozovic {\it et al.}, 
Phys. Rev. Lett. {\bf 59}, 2219 (1987).
\bibitem{warren} W. W. Warren {\it et al.}, 
Phys. Rev. Lett. {\bf 62}, 1193 (1989). 
\bibitem{puchkov} A. V. Puchkov, D. N. Basov, and T. Timusk, 
J. Phys. Condensed. Matter {\bf 8}, 10049 (1996).
\bibitem{loram1} J. W. Loram {\it et al.}, 
Proc. of 10th Anniversary HTS Workshop,
(World Scientific, Singapore, 1996).
\bibitem{tallon} J. L. Tallon {\it et al.}, 
Physica C {\bf 282-287}, 236 (1997).
\bibitem{harris}J. M. Harris {\it et al.},
Phys. Rev. B {\bf 54}, R15$\,$665 (1996).
\bibitem{ding} H. Ding {\it et al.},
Nature (London) {\bf 382}, 51 (1996).
\bibitem{hwang} H. Y. Hwang {\it et al.},
Phys. Rev. Lett. {\bf 72}, 2636 (1994).
\bibitem{chen1} X. K. Chen {\it et al.},
Phys. Rev. B {\bf 56}, R513 (1997).
\bibitem{emery} V. J. Emery and S. A. Kivelson, 
Nature (London) {\bf 374}, 434 (1995).
\bibitem{fisher} O. Fischer {\it et al.}, 
Physica C {\bf 282-287}, 315 (1997); Ch. Renner {\it et al.},
Phys. Rev. Lett. {\bf 80}, 149 (1998).
\bibitem{loram2} J. W. Loram {\it et al.} (unpublished).
\bibitem{tatiana} T. Startseva {\it et al.}, cond-mat9706145 (unpublished).
\bibitem{pines1} V. Barzykin and D. Pines, 
Phys. Rev. B {\bf 52}, 13585 (1995).
\bibitem{pines2} D. Pines, Physica C {\bf 282-287}, 273 (1997).
\bibitem{kimura} T. Kimura {\it et al.},
Physica C {\bf 192}, 247 (1992).
\bibitem{keimer} B. Keimer {\it et al.},
Phys. Rev. B {\bf 46}, 14034 (1992).
\bibitem{radaelli}P. G. Radaelli {\it et al.},
Phys. Rev. B {\bf 49}, 4163 (1994).
\bibitem{werder} D. J. Werder {\it et al.},
Physica C {\bf 160}, 411 (1989).
\bibitem{chen2} X. K. Chen, J. C. Irwin, R. Liang, and W. N. Hardy,
J. Supercond. {\bf 7}, 435 (1994); Physica C {\bf 227}, 113 (1994).
\bibitem{tom1} T. P. Devereaux, A. Virosztek, and A. Zawadowski
Phys. Rev. B {\bf 54}, 12523 (1996).
\bibitem{tom2} T. P. Devereaux and A. Kampf, cond-mat9711039 
(unpublished).
\bibitem{hackl} R. Hackl {\it et al.}, in
{\it Spectroscopic Studies of Superconductors}, edited by 
I. Bozovic and D. van der Marel (SPIE, Bellingham, WA 1996) p. 194.
%
\end{references}
\end{document}